\documentclass[twocolumn,preprintnumbers,secnumarabic,amsmath,amssymb,nofootinbib]{revtex4}
\usepackage{bm}

\usepackage{amsmath}
\usepackage{amsfonts}
\usepackage{amssymb}
\usepackage{graphicx}

\def\beq{\begin{equation}}
\def\eeq{\end{equation}}
\def\bea{\begin{eqnarray}}
\def\eea{\end{eqnarray}}

\def\CB {{\cal B}}

\newcommand\T{\rule{0pt}{2.6ex}}
\newcommand\B{\rule[-1.2ex]{0pt}{0pt}}

\begin{document}

\title{Inflating with Baryons}

\author{Daniel Baumann}
\email{dbaumann@ias.edu}
\affiliation{School of Natural Sciences, Institute for Advanced Study, Princeton, NJ 08540}
\author{Daniel Green}
\email{drgreen@ias.edu}
\affiliation{School of Natural Sciences, Institute for Advanced Study, Princeton, NJ 08540}

\begin{abstract}
We present a field theory solution to the eta problem.
By making the inflaton field the phase of a baryon of $SU(N_c)$ supersymmetric Yang-Mills theory we show that all operators that usually spoil the flatness of the inflationary potential are absent.  Our solution naturally generalizes to non-supersymmetric theories.
\bigskip

\end{abstract}

\maketitle

{\bf Introduction.}\
Inflation~\cite{Inflation} is remarkably successful as a phenomenological model of the early universe.
Besides solving the horizon and flatness problems of the standard big bang cosmology, inflation provides an elegant explanation for the observed fluctuations in the cosmic microwave background~\cite{Baumann:2008aq, Baumann:2009ds}. 
 However, despite considerable efforts, a concrete realization of inflation in a fundamental theory of particle physics remains elusive. 
The main challenge for achieving such a microscopic theory of inflation is the {\it eta problem}.
This refers to the problem of keeping the field which drives the inflationary expansion sufficiently light. If inflation is parameterized by a scalar field $\phi$ with potential energy
density
 $V(\phi)$, the quantitative requirement for prolonged slow-roll inflation is
\beq
\label{equ:eta}
\eta = M_{\rm pl}^2 \frac{V''}{V} \simeq \frac{m_\phi^2}{3H^2} \ll 1\ ,
\eeq
where the primes denote derivatives with respect to the field $\phi$.
Hence, inflation requires a hierarchy between the inflaton mass $m_\phi$ and the Hubble scale $H$.
It is 
difficult to protect this hierarchy against high-energy corrections.
Crucially,
even Planck-suppressed corrections
to the potential,
e.g.~dimension-six operators like $\Delta V \sim V(\phi) \frac{\phi^2}{M_{\rm pl}^2}$, generically destroy the condition (\ref{equ:eta}).

Our goal in this letter is to find a model of inflation for which the eta problem does not arise simply because there are no operators in the theory that can generate dangerous contributions to the inflaton mass.  It is well-known that  many contributions to the 
inflaton mass are absent when the 
low-energy effective action for the inflaton respects an approximate shift symmetry (see e.g.~\cite{Baumann:2010ys, McAllister:2007bg}).  Since shift symmetries can arise in field theory as a result of spontaneous symmetry breaking, it is natural to look for models where the inflaton is a pseudo-Nambu-Goldstone boson (PNGB)~\cite{NaturalInflation, ArkaniHamed:2003mz}.
However, quantum gravity effects are believed to break all global symmetries~\cite{Kallosh:1995hi}. Therefore, one cannot simply assume the presence of a shift symmetry without appealing to a UV completion.
 This UV-sensitivity has motivated the exploration of shift-symmetric large-field models in the context of string theory~\cite{Silverstein:2008sg}.  
Yet, to our knowledge, there is no proposed field theory mechanism that can explain the origin of this approximate symmetry in any viable model of inflation.
Of course, once a shift symmetry is {\it assumed} in the low-energy theory, the theory is radiatively stable~\cite{NaturalInflation, ArkaniHamed:2003mz, Baumann:2010ys}.
In contrast, in this letter we 
will aim for a more fundamental {\it explanation} for the protective symmetry.
We will present a class of models in which the approximate symmetry for the inflaton is an inevitable consequence of the particle content and 
the gauge symmetries of the theory.
 
Drawing inspiration from the stability of the proton, a natural approximate global symmetry to study is the baryon number symmetry of the Standard Model.  
Of course, this symmetry must be broken at high energies to explain the primordial baryon asymmetry in the universe. Moreover, 
it is necessarily broken by Planck-scale effects in a theory of quantum gravity.  However, experimental constraints on the lifetime of the proton rule out baryon number violation even through Planck-suppressed dimension-five operators.  This is not in tension with the Standard Model which  simply has no gauge-invariant baryon or lepton number violating operators with 
dimensions 
less than six.  In this letter we suggest that  the flatness of the inflaton potential can similarly be explained by a baryon number symmetry that cannot be broken by operators with dimensions less than seven.

\vskip 4pt
{\bf Supersymmetric~$SU(N_c)$.}\  As a concrete example of our proposal we will study ${\cal N}=1$ supersymmetric $SU(N_c)$ gauge theory~\cite{Terning:2006bq}.
The theory has $N_f$ {\it quarks} $q \equiv (q_i)_a$ and {\it anti-quarks} $\tilde q \equiv (\tilde q_{i})^a$. Often we will suppress the flavor indices $i$ and the gauge indices $a$.
We 
form gauge-invariant operators from $q$ and $\tilde q$ by contractions with the epsilon tensor or Kronecker 
deltas.
Using $\delta^a_b$, we construct the following {\it mesons}: $m_{ij} \equiv q_i \cdot \tilde q_j = \delta^{a}_b (q_i)_a (\tilde q_j)^{b}$ and $\hat m_{ij} \equiv q_i \cdot q^{\dag}_j = \delta^a_{b} (q_i)_a (q_{j}^\dagger)^{b}$.
Contractions
with the epsilon tensor 
result in {\it baryons} and {\it anti-baryons}~(for $N_f \geq N_c$):
\begin{align}
\CB_{i..k} &\equiv \epsilon^{a ..d}\, (q_{i})_a .. (q_{k})_d \ ,\\
  \tilde \CB_{i..k} &\equiv \epsilon_{a ..d} \,(\tilde q_{i})^a .. (\tilde q_{k})^d \ .
\end{align}
In the absence of a superpotential this theory has a $U(1)_B$ symmetry called the {\it baryon number symmetry}.
Under this symmetry the quarks have charge $+1$ ($q \to e^{i \alpha} q$) and the anti-quarks have charge $-1$ ($\tilde q \to e^{- i \alpha} \tilde q$).   The mesons are invariant under the symmetry, while the baryons are charged: e.g.~$\CB_{i..k}  \to e^{i N_c \alpha} \CB_{i..k} $.  From here on, we will be a bit careless about the flavor indices unless this could be a source of confusion.  It will ultimately be one specific baryon that plays the important role. We will call this baryon simply  $\CB$ (or $\tilde \CB$).  The phase of the baryon will become the inflaton field. Its potential will be protected by the $U(1)_B$ symmetry.

\vskip 4pt
{\bf $U(1)_B$ and gravity.}\ 
Let us 
explain why $U(1)_B$ is a good symmetry even when coupled to gravity, despite the common lore that a theory of quantum gravity breaks all global symmetries~\cite{Kallosh:1995hi}.  To understand this we 
consider which operators we could write down that would break the symmetry.  For simplicity, we take $N_f > 3 N_c$, in which case the theory is weakly coupled at low energies and we can treat all fields as free~\cite{Terning:2006bq}.  The dimension of the baryon $\CB$ is then $N_c$.  Hence, for $N_c \geq 3$ there 
are literally no relevant operators in the theory that could break baryon number.   This simple fact lies at the heart of our solution to the eta problem.  As we will show, it leads to a parametric suppression of K\"ahler corrections. 
This is a dramatic advantage of our approach since, in general, corrections to the K\"ahler potential are extremely dangerous and often poorly controlled.

\vskip 4pt
{\bf A simple model.}\ 
We now write down a simple model 
of supersymmetric hybrid inflation that realizes our basic idea.
We will start by assuming that there is a singlet spurion field $X$ whose non-zero F-term drives inflation, $F_X = \mu^2$.  
We will further assume that baryon number is spontaneously broken at a scale~$f > \mu$.  
We identify the phase of $\CB$ with the inflaton field $\phi \equiv \sqrt{2 N_c} f \theta$,
i.e.~we define $\CB \sim f^{N_c} e^{N_c i\theta}$.
The inflationary regime will be at small field values, $\phi < f$.

The 
K\"ahler potential for the quarks and the spurion has the canonical form
\beq
K = q_i\cdot q^\dag_i +  \tilde q_i \cdot \tilde q^\dag_i + X X^\dag\ .
\eeq
To this we add a superpotential that spontaneously breaks the $U(1)_B$ symmetry and gives rise to inflation 
\beq  
\label{equ:WW}
W = S^{mn} (q_{m} \cdot \tilde q_{n} - f^2 \delta_{mn}) -  \mu^2 X \ ,
\eeq
where $S^{mn}$ are singlets that will be integrated out. 
The flavor indices  $m, n$ run over the first $N_c$ flavors, while the gauge indices of $(q_m)_a$ and $(\tilde q_n)^{b}$ are contracted.  
This model has the basic features we require:
The first term in the superpotential induces a symmetry breaking vacuum expectation value (vev) for the quarks $(q_{m})_a \sim f e^{i \theta} \delta_{m,a}$ 
(from the F-term equations for $S^{mn}$), while the second term leads to a constant vacuum energy, $V_0 \approx |F_X|^2 = \mu^4$. Finally, the phase of the baryon $\CB$ is a Goldstone mode.  Importantly, it should be clear that the $U(1)_B$ symmetry is only broken by {\it irrelevant} couplings in the superpotential.  Therefore, $U(1)_B$ is necessarily a good global symmetry at low energies, even if it is broken badly at the Planck scale.  

\vskip 4pt
{\bf Absence of the eta problem.}\ 
The eta problem usually arises because Planck-suppressed operators of dimension up to six induce a large inflaton mass. We now explain 
in more detail
why this does not happen for our baryon inflation scenario.

First, we note that the K\"ahler potential can only give rise to mass terms for the PNGB through operators of the form 
\beq
\Delta K \sim \frac{(\CB + \CB^\dag)}{\Lambda^{N_c}}X^{\dag} X\ ,
\eeq  
where $\Lambda \leq M_{\rm pl}$ is the cutoff of the low-energy  effective theory.
The associated correction to the inflaton potential takes the form 
\beq
\Delta V \sim \mu^4 {f^{N_c} \over \Lambda^{N_c}} \cos\bigl(N_c \theta \bigr)\ ,
\eeq  
and the corresponding contribution to $\eta$ is  
\beq
\Delta \eta \sim \Bigl({f \over \Lambda} \Bigr)^{N_c - 2} \cdot \frac{M_{\rm pl}^2}{\Lambda^2}\ .
\eeq
For given $\frac{f}{M_{\rm pl}}$ and $\frac{\Lambda}{M_{\rm pl}}$ we can always find a $N_c \geq 3$ such that $\eta$ is very small.
By construction, we can therefore ignore corrections to the K\"ahler potential entirely.

Furthermore, 
there are only a handful of operators that 
can be added to the superpotential given the particle content and symmetries of the theory:  

- A potentially dangerous operator in the superpotential is 
\beq
\Delta W \sim   { (\CB + \tilde \CB) \over \Lambda^{N_c -2}} X\ ,
\eeq
which contributes the following correction to the eta parameter,
\beq
\label{equ:etaWX}
\Delta \eta \sim \Bigl( \frac{f}{\Lambda} \Bigr)^{N_c-4} \cdot \frac{f^2}{\mu^2} \cdot \frac{M_{\rm pl}^2}{\Lambda^2}\ .
\eeq
For sufficiently large $N_c \ge 4$
this will be a small correction, if we take into account the typical hierarchies of scales:
\vskip -5pt
 \begin{figure}[h!]
   \centering
       \includegraphics[width=.38\textwidth]{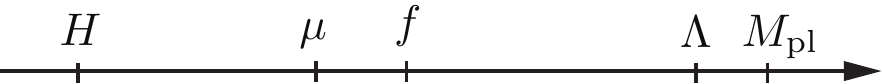}
   \label{fig:scales}
\end{figure}
\vskip -5pt

- Next, we consider the effect of adding $\CB$ (or $\tilde \CB$) directly to the superpotential (\ref{equ:WW}) 
\beq
\Delta W \sim \frac{\CB}{\Lambda^{N_c-3}}\ .
\eeq
This causes $S_{mn}$ to get a vev: 
\beq
S_{mn} \sim e^{(N_c-2)i \theta} \frac{f^{N_c -2}}{ \Lambda^{N_c -3}}\ ,
\eeq
and induces a mixing between $S_{mn}$ and the inflaton. This can 
give rise to a mass term for $\theta$ through the following coupling in the K\"ahler potential: $\Delta K \sim \Lambda^{-1}(S_{mn} + S_{mn}^\dag) X^{\dag} X$.
The resulting contribution to eta is 
\beq
\label{equ:EtaSij}
\Delta \eta \sim \Bigl(\frac{f}{\Lambda}\Bigr)^{N_c - 4} \cdot \frac{M_{\rm pl}^2}{\Lambda^2} \ .
\eeq
We note that this is always smaller than $\Delta \eta$ in (\ref{equ:etaWX}), so it doesn't give a new constraint.

\vskip 4pt
{\bf Graceful exit problem.}\
As is typical in small-field models, additional {\it waterfall fields}  are required to end inflation and reduce the effective 
vacuum energy to zero. 
To construct such a model of {\it hybrid inflation}~\cite{Hybrid} we introduce additional singlets $\psi$ and $\tilde \psi$ which couple to $\CB$ and $X$.
The superpotential becomes
\beq 
W = S^{mn} (q_{m} \cdot \tilde q_{n} - f^2 \delta_{mn}) + \lambda\, \frac{(\CB + \tilde \CB)}{\Lambda^{N_c-1}} \, \psi \tilde \psi + (y^2 \psi^2 - \mu^2) X\ ,
\eeq
where $\lambda$ and  $y$ are dimensionless coupling constants.
The phase of $\CB$ is 
now a PNGB.
The masses of the $\psi$ and $\tilde \psi$ fields receive a contribution that depends on the inflaton vev: $\Delta m_{\psi} = \hat \lambda f \cos(N_c \theta)$, where $ \hat \lambda \equiv \lambda {f^{N_c-1} \over \Lambda^{N_c -1}} $.  
For the conventional hybrid mechanism we require that the waterfall mass during inflation 
(i.e.~for small $\theta$) exceeds the Hubble scale $\hat \lambda f > H \sim \mu^2 M_{\rm pl}^{-1}$, or
\beq
\label{equ:p}
\lambda \ >\ \Bigl(   \frac{f^{N_c-4}}{\Lambda^{N_c-4}}  \cdot \frac{f^2}{\mu^2} \cdot \frac{M_{\rm pl}^2}{\Lambda^2} \Bigr)^{-1} \cdot \frac{\Lambda^2}{f^2} \cdot \frac{M_{\rm pl}}{\Lambda} \ .
\eeq
Here, we have written the first factor as the inverse of $\Delta \eta$ in (\ref{equ:etaWX}).
The inequality in (\ref{equ:p}) illustrates a significant problem: 
the requirement $\Delta \eta < 1$ in (\ref{equ:etaWX}) implies $\lambda \gg 1$.

In a sense, we have become victims of our own success: The only $U(1)_B$ breaking couplings to the singlets $\psi$ and $\tilde \psi$ are irrelevant couplings. Hence, for the same reason that the inflaton potential is very flat, the couplings of the inflaton to the waterfall fields are very weak. In fact, 
the couplings are too weak to lead to the traditional hybrid inflation scenario.

\begin{table}[h!]
\begin{tabular}{| l | l |c |}
\hline
\hspace{0.15cm}  \T \B   &\hspace{.7cm}   & \hskip 4pt $U(1)_B$ \hskip 4pt  \\
\hline
\hskip 2pt quarks \T  & \hskip 2pt $(q_i)_a$ & \hskip 2pt $+1$ \hskip 4pt \\
  \T \B & \hskip 2pt $(\tilde q_i)^{a}$ & \hskip 2pt $-1$ \hskip 4pt \\
\hskip 2pt mesons \T \B &   \hskip 2pt  $m_{ij} = q_i \cdot \tilde q_j$  &
$0$ \hskip 2pt\\
 \T \B &   \hskip 2pt  $\hat m_{ij} = q_i \cdot q_j^\dag$ &
$0$ \hskip 2pt\\
\hskip 2pt  baryons \T \B &  \hskip 2pt   $\CB_{i..k} = \epsilon^{a ..d}\, (q_{i})_a .. (q_{k})_d$ \hskip 4pt  & $+N_c$  \hskip 2pt\\
\T  \B &  \hskip 2pt   $\tilde \CB_{i..k} = \epsilon_{a .. d}\, (\tilde q_{i})^{a} .. (\tilde q_{k})^{d}$ \hskip 4pt  & $-N_c$  \hskip 2pt\\
\hskip 2pt spurion  \T \B &  \hskip 2pt   $X$ \hskip 4pt  & $0$  \hskip 2pt\\
\hskip 2pt mediator \T \B &  \hskip 2pt   $h_{ab}$ \hskip 4pt  & $0$  \hskip 2pt\\
\hskip 2pt waterfall \B &  \hskip 2pt   $\psi$ ($\tilde \psi$) \hskip 4pt  & $0$  \hskip 2pt\\
\hline
\end{tabular}
\caption{Matter content and symmetries.}
\end{table}

\vskip 4pt
{\bf Improved waterfall coupling.}\
The following solution to the above problem suggests itself:
we introduce larger representations (rather than singlets) to mediate $U(1)_B$ breaking effects to the waterfall fields via direct couplings to quarks. After integrating out the massive mediator fields this leads to marginal (rather than irrelevant) couplings between the inflaton and the waterfall fields.

Let us give 
an explicit
example to see how this 
works in practice.
For concreteness, we will present the special case of $N_c = 5$.  To the model above, we will add a $\mathbf{10}$, denoted by $h \equiv h_{ab}$ (we also add a spectator $\overline{\mathbf{10}}$ to cancel anomalies).  We break up the quark flavors into $q_m$ for $m \leq 5$ and $Q_k$ for $k > 5$.  
The dynamics of these fields is governed by the following superpotential
\bea
\label{equ:super}
W &=& S^{mn} (q_m\cdot  \tilde{q}_n - f^2 \delta_{mn})  \nonumber \\
&& +\  ( y^2 \psi^2 -\mu^2) X  + m\, \psi \tilde \psi  + \lambda\, q_1 \cdot h \cdot h \\
&& +\  (\psi\, q_3 + \tilde{q}_2\cdot h ) \cdot \tilde{Q}_6 + (\tilde \psi\, q_5 + \tilde{q}_4 \cdot h ) \cdot  \tilde{Q}_7\ . \nonumber
\eea
The $\mathbf{5}$-$\mathbf{10}$-$\mathbf{10}$ coupling $q_1\cdot h \cdot h = \epsilon^{abcde} (q_1)_a h_{bc} h_{de}$ breaks the $U(1)_B$ symmetry.
The $\mathbf{1}$-$\mathbf{5}$-$\mathbf{\overline{5}}$ and $\mathbf{\overline{5}}$-$\mathbf{10}$-$\mathbf{\overline{5}}$ couplings in the last line of (\ref{equ:super}) will mediate this to the waterfall field~$\psi$.  
At the scale $f$,  the quarks $q$ and $\tilde q$ get vevs which give masses to 
the fields $\tilde{Q}$ and $h$.   We will assume that $m < f$ so that we can treat $X$, $\psi$ and $\tilde \psi$ as the light fields.  Integrating out $\tilde{Q}$ and $h$ then leads to the required structure for the effective couplings of the light fields.

First, we use the equations of motion to give vevs to $N_c$ flavors of the quarks:
\beq
(q_{m})_a = f c\, e^{i \theta} \delta_{m,a}\quad {\rm and} \quad
(\tilde{q}_{n})^b = f c^{-1} e^{-i \theta} \delta_{n}^b \ ,
\eeq 
where $c$ is a pseudo-modulus. 
As we will discuss below, a stabilizing potential is generated  for $c$ by loop and/or K\"ahler corrections. For now we assume that 
 it gets stabilized at $c \sim 1$.
The equations of motion for $\tilde{Q}_6$ and $\tilde{Q}_7$ then lead to vevs for the 
mediator fields: 
\beq
h_{2 3} = \psi \, c^{2}\, e^{2i  \theta} \quad {\rm and} \quad
h_{45} = \tilde \psi  \, c^{2}\, e^{2i \theta}\ .
\eeq
Hence, after integrating out the massive fields, the effective superpotential becomes
\beq
\label{equ:Weff}
W_{\rm eff} =m \left(1 + d \, e^{ 5i\theta}   \right) \psi \tilde \psi  +  X \left( y^2 \psi^2- \mu^2\right)\ ,
\eeq
where
\beq
\label{equ:inequality}
d \equiv \frac{f}{m}\, \lambda c^5 \ .
\eeq
The structure of (\ref{equ:Weff}) is exactly the same as that of the superpotential previously studied by us in Ref.~\cite{Baumann:2010ys} (see also \cite{ArkaniHamed:2003mz}).
For $d \sim 1$ the mass for $\psi$ is roughly $m$ for $\theta \sim 0$ but vanishes 
near 
$\theta \sim \pi / 5$ 
and becomes tachyonic for $\theta \ge \pi/5$.  This is a standard waterfall potential that will stabilize the field at $\psi = 0$ during inflation provided that $m^2 > y^2 \mu^2$.  Inflation ends when $\theta\sim \pi /5$ and $\psi$ acquires a vev.  

The model 
in (\ref{equ:super}) is still somewhat incomplete. Specifically, in deriving the effective superpotential (\ref{equ:Weff}) we fixed the vevs of the massless scalars $X$ and $c$.  These fields are pseudo-moduli and do not stay massless once supersymmetry 
(SUSY) is broken.  For example, $c$ is the superpartner of a Goldstone boson, the inflaton $\theta$, and its mass is only protected by SUSY.  When SUSY is broken, $c$ is not protected from K\"ahler corrections and generically receives at least a Hubble scale mass.  Furthermore, if any pseudo-modulus is coupled to additional massive fields in the superpotential, a potential for the modulus will be generated through loops.  It then remains only to show that this potential does not contain runaways.  A comprehensive treatment of this question and various mechanisms to stabilize pseudo-moduli appeared 
in Ref.~\cite{Intriligator:2008fe}.  It is straightforward to apply these techniques to $c$ and $X$ such that they are stabilized near one and zero, respectively.  As we will discuss next, these effects rarely contribute to the potential for the inflaton.    

\vskip 4pt
{\bf Revisiting the eta problem.}\ 
Since we have introduced the mediator field $h_{ab}$ there are now additional gauge-invariant operators that we can write down. 
We need to check that these new operators don't reintroduce the eta problem through the backdoor.

What makes many of these operators harmless is the fact that the waterfall field doesn't get a vev during inflation.  Hence, any operators involving $\psi$ and $h$ coupled to $X^{\dag}X$ in the K\"ahler potential do not modify the inflaton potential.  However, superpotential couplings may modify the potential both at tree level and at one-loop. 
As we show next, these
can quite easily be made small enough to avoid the eta problem.

We 
will systematically characterize 
all possible modifications to the superpotential and the K\"ahler potential that could alter the potential for the inflaton or the waterfall fields.  It will be convenient to break up the fields into three groups: the inflaton sector ($q,\tilde{q}$), the vacuum energy ($X, S_{mn}$) and everything else ($Y \equiv \{h, \psi, \tilde \psi, Q, \tilde Q\}$).  The possible deformations are then classified by the factors of $Y$ in a given operator:

- The case with no factors of $Y$ is identical to the model without any waterfall fields (\ref{equ:WW}).  As we discussed above, the K\"ahler corrections give a highly suppressed contribution to the inflaton mass.  In addition, the most significant superpotential deformations $\CB / \Lambda^{N_c -3}$ and $ \CB X / \Lambda^{N_c -2}$ are suppressed for $N_c > 4$, cf.~(\ref{equ:etaWX}) and (\ref{equ:EtaSij}). In particular, 
their contributions to $\eta$ are small for our explicit example with $N_c =5$.

- Linear terms in $Y$ are potentially dangerous as they can completely change the vacuum structure of the potential.  For example, the addition of operators like $\mu^2 \psi$ to the superpotential will give 
unwanted vevs (or F-terms) to some fields.  Operators of the form $q \cdot h \cdot q$ will have the same effect.  Rather than putting bounds on the coefficients of these operators, it is easier to introduce a $\mathbb{Z}_2$ symmetry under which $Y \to - Y$.  One can check that the superpotential (\ref{equ:super}) is invariant under this symmetry.  The presence of this symmetry would also explain the form of the potential that distinguishes $q, \tilde{q}$ from $Q, \tilde Q$.

- Quadratic terms in $Y$ do not modify the potential for the inflaton at tree level, although they may contribute at one-loop.  Generic terms quadratic in $Y$ are independent of the phase of the inflaton.  This should be evident from our difficulty in constructing a suitable mass for the waterfall fields.  For such fields, loop corrections will contribute only to the Coleman-Weinberg potential for the pseudo-moduli (e.g.~$c$ and $X$).  For the few fields, like $\psi$, whose mass depends on both $\phi$ 
and $X$, the Coleman-Weinberg potential will contribute to the mass of $\phi$.  
In particular, the
potential associated with the couplings in (\ref{equ:Weff}) is
\beq
V_{\rm cw}(\phi) = \mu^4 \frac{y^4}{4\pi^2} \, \log \Bigl(\frac{m}{\sigma} \cos(\phi/f) \Bigr)\ ,
\eeq
where $\sigma$ is the renormalization scale. Keeping the contribution to the eta parameter small imposes a constraint on the coupling constant,  $y^2 \ll \frac{f}{M_{\rm pl}}$. 
The small size of the coupling is technically natural and quite common in SUSY hybrid models~\cite{ArkaniHamed:2003mz}. Furthermore, it seems conceivable that the small value of $y$ can be explained using non-perturbative physics~\cite{Dine:2006gm}.

- Quadratic terms in $Y$ can also modify the 
waterfall potential.
Additional operators in the superpotential
provide
the most stringent constraints.  For example, including $ \xi\, q_2 \cdot h  \cdot Q_8$ in the superpotential gives an additional mass to the waterfall fields $m_{\rm add}^2 \sim \xi^2 f^2$.  An acceptable waterfall potential implies that the effective superpotential has the following hierarchy: $ m^2 > y^2 \mu^2 > m^2_{\rm add}$.  We should stress that this hierarchy is common to all models of hybrid inflation and in particular is completely independent of our specific solution to the eta problem. 
Quadratic operators of the form $\varsigma\, Y^{\dag} Y \frac{X^{\dag} X}{ \Lambda^2}$ in the K\"ahler potential 
also contribute additional masses to the waterfall fields.  However, these contributions can be ignored provided that $\varsigma < y^2 \Lambda^2 / \mu^2$.  Then the hierarchy of masses we require involves only terms in the superpotential and therefore is 
analogous
to the hierarchy of Yukawa couplings in the MSSM.

\vskip 4pt
{\bf Discussion and Outlook}.\ 
In this letter we have proposed a new solution to the eta problem based on
the $U(1)_B$ baryon symmetry of $SU(N_c)$ gauge theory. The inflaton field was identified with the phase of a baryon field. It is then easy to see that for $N_c \ge 3$ there is simply no relevant operator in the theory that could break baryon number.
This gives a powerful way to protect the inflaton mass from K\"ahler corrections which otherwise are notoriously hard to characterize.

The reader should rightly be suspicious of any claims of a natural solution to the eta problem and suspect that hidden corrections have been missed or that the problem has simply been moved to a different sector.
As an existence proof we therefore constructed a detailed model with all the required features. We explained why coupling the baryon phase to waterfall fields requires some care and presented a viable hybrid mechanism in a model with $SU(5)$ symmetry.
In the most obvious extensions of the $SU(5)$ model to arbitrary $N_c$, the waterfall fields obtain 
masses 
through at least one irrelevant operator.  Unlike the $SU(5)$ model, such terms will imply a lower bound 
on the hierarchy ${f \over \Lambda} \gtrsim \bigl(\frac{H}{f}\bigr)^{1/n}$, for some integer $n$. 
These minimal models are still viable, but they don't lead to arbitrarily large suppressions in the $N_c \to \infty$ limit.  One could hope to 
realize this parametric suppression by altering the waterfall potential, but we leave a systematic exploration of this question to future work~\cite{progress}.

We stress that our idea is 
more general than the specific model we have presented.
It would be very interesting to find other examples and characterize the space of possible effective Lagrangians.
Furthermore, it may be instructive to study possible UV completions of the structures we introduced in this letter. 
Since all elements of our theory have direct counterparts in the Standard Model we don't see a fundamental obstruction to such an endeavor.  Even if the UV completion contains additional matter, charged or neutral, the above analysis can be repeated
by including them as additional $Y$~fields.  
Finally, 
we remark that we don't believe that 
supersymmetry played an important role in our mechanism.
While we found supersymmetry a convenient framework to write down a radiatively stable model, our idea to use the high dimensionality of symmetry breaking operators in itself seems to have a wider range of applicability.

\vskip 10pt
{\it Acknowledgements.}\
We thank N.~Craig, S.~Kachru, L.~McAllister, and E.~Silverstein for comments on a draft of the manuscript and A.~Dymarsky, Z.~Komargodski and J.~Heckman for helpful discussions.
The research of D.B.~is supported by the National Science Foundation under grant number AST-0855425. The research of D.G.~is supported by the Department of Energy under grant number DE-FG02-90ER40542.

\vfil

\begingroup\raggedright\endgroup

\end{document}